\documentclass[aip,amsmath,twocolumn,amssymb,seceqn,10pt]{revtex4-1}
\usepackage{graphicx}
\usepackage{graphics}
\usepackage{epsfig}
\usepackage{dcolumn}
\usepackage{bm}
\usepackage{makeidx}
\usepackage{subfigure}
\usepackage{color}
\usepackage{longtable}

\begin{document}

\title{Part of a collection of reviews on antiferromagnetic spintronics. Antiferromagnetic dynamics, spin-texures, and nanostructures}
\author{O.~Gomonay}
\affiliation{Institut f\"ur Physik, Johannes Gutenberg Universit\"at Mainz, D-55099 Mainz, Germany}
\affiliation{National Technical University of Ukraine "KPI", 03056 Kyiv, Ukraine}
\author{V.~Baltz}
\affiliation{SPINTEC, Univ. Grenoble Alpes / CNRS / INAC-CEA, F-38000 Grenoble, France}
\author{A.~Brataas}
\affiliation{Department of Physics, Norwegian University of Science and Technology, NO-7491 Trondheim, Norway}
\author{Y.~Tserkovnyak}
\affiliation{Department of Physics and Astronomy, University of California, Los Angeles, California 90095, USA}

\begin{abstract}
Antiferromagnets as active elements of spintronics can be faster than their ferromagnetic counterparts and more robust to magnetic noise. Owing to the strongly exchange-coupled magnetic sublattice structure, antiferromagnetic order parameter dynamics are qualitatively different and thus capable of engendering novel device functionalities. In this review, we discuss antiferromagnetic textures -- nanoparticles, domain walls, and skyrmions, -- under the action of different spin torques. We contrast the antiferromagnetic and ferromagnetic dynamics, with a focus on the features that can be relevant for applications. 

\end{abstract}

\maketitle

%

\section{Introduction}
An important drive in antiferromagnetic spintronics is the active manipulation of the antiferromagnetic state and its magnetic spin-textures via spin and charge currents. 
For example, a direct electrical re-orientation of antiferromagnetic domains by the  current-induced N\'eel spin-orbit torque has been recently demonstrated. \cite{Wadley2016}
However, the dynamics of this process and other excitation processes, and of the antiferromagnetic textures
remain to be fully understood and exploited. 

A key aspect to understand is the fundamental qualitative difference in the dynamics of antiferromagnets (AFs) and ferromagnets (FMs). The exchange coupling between the sublattices in an AF brings about a more complex and in general faster dynamics as compared to FMs. Hence, a large part of the intuitive thinking arising from the ferromagnetic case has to be re-examined in AFs. 
In addition, the possibilities of different types of current-induced spin torques are also expanded beyond the ferromagnetic case. 

Excitations of antiferromagnetic structures -- domain walls, skyrmions, and spin-waves -- also demonstrate a number of peculiar features that can endow new functionalities to antiferromagnetic-based spintronic devices, as compared to ferromagnetic ones. Their current-induced dynamics is a very active research area of antiferromagnetic spintronics. Below we focus on emphasising the stark differences between AFs and FMs, and the new possibilities that AFs offer for spintronics.

\section{General features of antiferromagnetic dynamics}
AFs are materials with long-range magnetic order and a vanishingly small or zero macroscopic magnetization. The simplest AFs are collinear consisting of  anti-aligned magnetic moments $\mathbf{M}_1$ and $\mathbf{M}_2$ belonging to two magnetic sublattices.  The description of the antiferromagnetic dynamics is based on equations of motion for the magnetic sublattice moments similar to those that are used for ferromagnetic  systems but coupled via an inter-sublattice exchange field, $H_{ex}$.\cite{Kittel1951,Keffer1952}
Although it might be natural to think that such a system behaves like two interpenetrating FMs, the dynamics of AFs is richer and more complex than the dynamics of FMs. 

In FMs, the homogeneous (zero-momentum) resonance excitation (FMR) is determined by the energy scales of the anisotropy fields, $H_{an}$ (which include both magnetocrystaline and dipolar terms), whose order of magnitude is typically GHz. Any involvement of exchange energies in FMs requires finite momentum magnon excitations that demagnetize the system. On the other hand, in AFs, in the presence of crystalline anisotropies,
the inter-sublattice exchange allows for optical-mode homogeneous excitations that involve $H_{ex}$. More specifically, these antiferromagnetic resonance (AFR) excitations have typically an enhancement factor of $\sqrt{H_{ex}/H_{an}}$ relative to the FMR, which brings them to THz frequencies. \cite{Kittel1951,Keffer1952,Ross2015}
In addition, a relatively high value of the spin-flop field \cite{Hagiwara1999} compared to its analogue in FMs, the coercive field, also has this exchange enhancement factor.



\begin{figure}[t]%
	\includegraphics[width=.8\columnwidth]{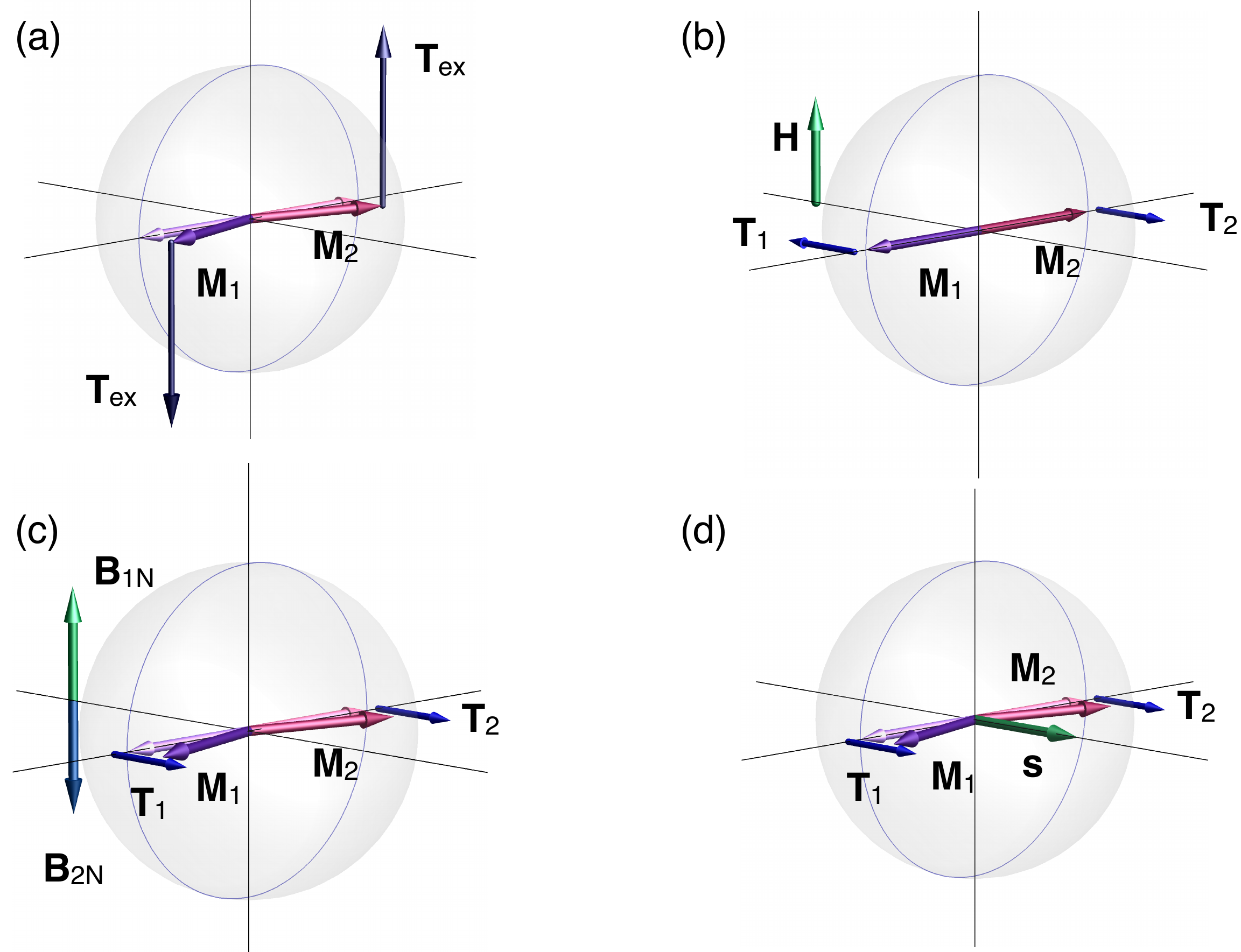}
	\caption{(Color online) Torques and dynamics of AF. (a) Sublattice magnetizations $\mathbf{M}_1$ and $\mathbf{M}_2$ are antiparallel in equilibrium. Any excitation followed by a small tilt of $\mathbf{M}_1$ and $\mathbf{M}_2$ triggers oppositely directed large torques $\mathbf{T}_\mathrm{ex}$ 
		which induce fast rotation of the magnetic moments. Oppositely, due to the tilting, any rotation of $\mathbf{M}_1$ and $\mathbf{M}_2$ is associated with appearence of nonzero dynamic magnetization. (b) Magnetic field $\mathbf{H}$ generates the antiparallel torques $\mathbf{T}_\mathrm{1/2}$ which compensate each other and, thus, hamper magnetization dynamics.
The staggered N\'eel field $\mathbf{B}_\mathrm{1N/2N}$ (c) and the current with spin polarization $\mathbf{s}$ (d) generate the parallel torques $\mathbf{T}_\mathrm{1/2}$ which cant sublattice magnetizations $\mathbf{M}_{1/2}$ forward and, thus, create internal exchange torques $\mathbf{T}_\mathrm{ex}$ which cause rotation of magnetic sublattices. 
	}
	\label{fig_AF_torques}
\end{figure}

Exchange enhancement also appears in the dynamics of antiferromagnetic spin textures.   In a FM,  the velocity of a domain wall is limited by the Walker breakdown. On the other hand, in an AF, the limit is set  by the magnon velocity, \cite{Gomonay2016, Shiino2016} which due to the
strong exchange enhancement, is much larger than the typical magnon velocity in a FM. In addition, the internal exchange torques in an AF are several orders of magnitude larger than any driving torque, which leads to a stiff domain wall with low effective mass and shifts the point of the Walker breakdown to an unreachable driving field value. \cite{Gomonay2016, Selzer2016} Recently, a high domain wall velocity (up to 2000 m/sec) which is one order of magnitude larger than the limiting domain wall velocity in FMs, was detected at the compensation point of a ferrimagnet. \cite{Kim2017}

A further distinct aspect in AFs, besides the exchange enhancement effect, is the issue of which torques are more efficient in exciting and manipulating the state of the AF and its magnetic textures. 
In a FM, one can use, apart from a uniform magnetic field, the spin-transfer or spin-orbit torques. These spin torques can be field-like or antidamping-like. For example, the switching due to the antidamping-like spin-transfer torque takes place when the internal damping is compensated for one direction of the applied spin current. After switching, however, the same current stabilizes the reoriented magnetization by  adding to its damping in the stable direction. This leads to a well-controled switching that is robust with respect to fluctuations of the driving current parameters. 

In an AF, a uniform static field generates torques that have opposite sign on the opposite sublattices (Fig.1(b)). These torques almost compensate each other at the macroscopic scale and are relatively inefficient  in switching AFs. On the other hand, the antidamping-like spin transfer torque in an AF due to a uniformly polarized injected spin-current is an efficient generator of oscillations of the staggered magnetization. In a  configuration  where the polarization of the spin current is perpendicular to the N\'eel vector (Fig.1(d)), it generates parallel torques on the opposite magnetic sublattice. 
Above a threshold these torques can induce a stable precession of the staggered magnetization within the plane perpendicular to the spin current polarization. \cite{Gomonay2008,Cheng2015} Similar effect can be induced by the spin Hall effect in a bilayer consisting of an AF and a heavy-metal. \cite{Zarzuela2017}

If the spin polarization is parallel to the N\'eel vector, the spin current can induce an instability of the initial state. However, unlike in a FM, the spin polarization in an AF is always antiparallel to one of the magnetic sublattices. This means that in this configuration the spin transfer torque in an AF always competes with the internal damping on one of the spin -sublattices and  the staggered magnetization tends to rotate towards the plane perpendicular to the spin polarization. The final state in this case also corresponds to a stable precession.

Thus, while the FM driven by a magnetic field or a spin-polarized current tends to switch between the different static states, an AF is a natural spin-torque oscillator. \cite{Gomonay2014,Cheng2014c} Typical frequency of such an oscillator  
scales with the AFR frequency and falls into the THz range. It can be tuned by the magnitude of the injected spin-current. This feature of AFs is attractive for applications. In particular, as the precession of staggered magnetization creates a nonzero dynamic magnetization (Fig.1(a)), such spin torque oscillator can pump spin current into the neighboring nonmagnetic layer, \cite{Gomonay2014,Cheng2014c} similar to a FM. Spin pumping can also occur due to rotation of staggered magnetization in the course of a spin superfluid transport. \cite{Takei2014b} 

In addition, the time-dependent component of the dynamic magnetization can be a source of a THz signal either through the direct emission of electromagnetic waves or due to pumping of THz spin current into the electrodes.  
The additional flexibility of this type of the device can be achieved by 
the modification of the internal damping due to, e.g., a feedback from the nearby electrodes. 
This allows to control the angle between the staggered magnetization and spin polarization by the magnitude of the spin current.\cite{Cheng2016} Similar effects of  THz emission can be also observed in ferrimagnetic materials. \cite{Awari2016} These materials combine the advantages of strong exchange coupling between the magnetic sublattices specific to AFs with the nonzero macroscopic magnetization which facilitates easy detection of the magnetic dynamics.  
 

There is also a distinct type of torque that is not relevant in FMs. This is the N\'eel spin-orbit torque, which was proposed \cite{Zelezny2014}  in 2014 and recently experimentally observed. \cite{Wadley2016} The torque can be generated in certain AFs where a staggered field $\mathbf{B}_\mathrm{N}$ with opposite signs on opposite sublattices is induced by a global uniform current. The staggered field in AFs plays an analogous role to the uniform magnetic  field in a FM. The staggered field generates torques on the antiferromagnetic spin-sublattices which sum up on the macsocopic level (Fig.1(c)). They rotate the staggered magnetization and vanish when the staggered magnetization is parallel to $\mathbf{B}_\mathrm{N}$. Thus, the N\'eel spin-orbit torque induces fast and deterministic switching in an AF.  
 
 The concept of the magnetic sublattices and corresponding torques is applicable to macrospin models of AFs. However, electrical current allows also for the manipulation of antiferromagnetic nanostructures (see Fig.2(a)) which include a small number of magnetic atoms. For example, with the use of a spin-polarized tip,  heating combined potentially with the spin transfer torque was applied directly to a particular atom in a one-dimensional chain of antiferromagnetically ordered Fe moments.\cite{Loth2012} Reorientation of the whole magnetic nanostructure was then mediated by strong exchange coupling between the neighboring magnetic atoms. 

\begin{figure}[t]%
	\includegraphics[width=.8\columnwidth]{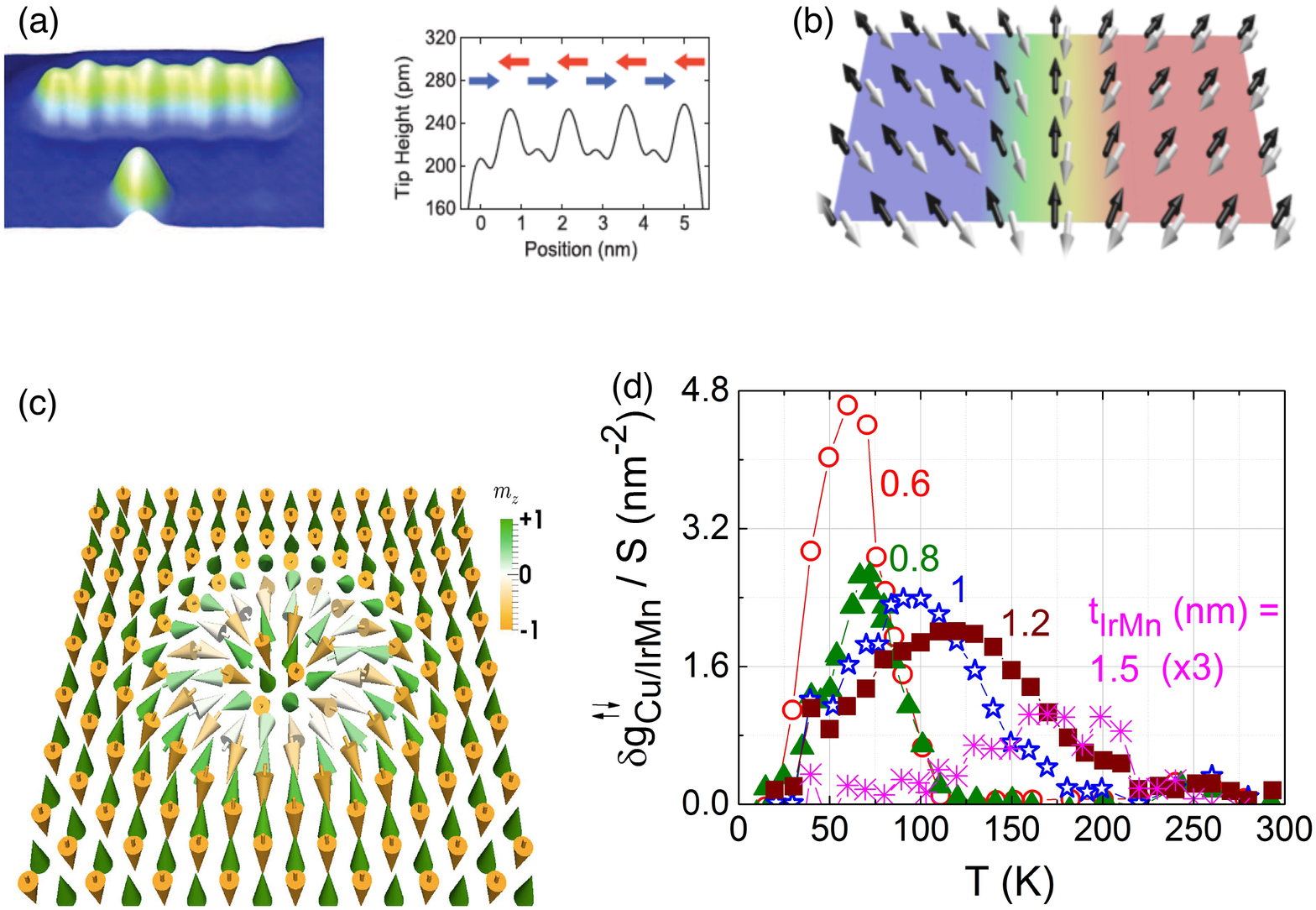}
	\caption{(Color online) Antiferromagnetic bits (a)  Spin polarized STM image of a nanostructure consisting of antiferromagnetically coupled eight Fe atoms assembled on a Cu$_2$N overlayer on Cu(100). From Ref.\onlinecite{Loth2012}. Illustration of 90$^\circ$ AF domain wall (b) and AF skyrmion spin texture (c), from Ref.\onlinecite{Zhang2015b}. Sublattice magnetizations $\mathbf{M}_1$ and $\mathbf{M}_2$ (shown in different colors) smoothly rotate in space keeping mutual antiparallel orientation.  (d) Spin transmission. Enhanced spin transmission occurs during the magnetic phase transition of AFs illustrating that a fluctuating magnetic order allows more spins to pass through an interface. Data deduced from spin pumping experiments with NiFe/Cu/IrMn metallic multilayers in which the IrMn layer thickness ($t_\mathrm{IrMn}$) is varied. Thinner layers display smaller critical temperature for the magnetic phase transition as a result of finite size effects. Adapted from Ref. \onlinecite{Frangou2016}.
	}
	\label{fig_AF_structures}
\end{figure}

\section{Domain walls and textures}

Application of antiferromagnets in spintronics is not limited to uniform systems. Antiferromagnetic textures, like  domain walls and skyrmions, could play an important role in information coding and have intriguing physical properties.
 
\emph{Domain walls} give the simplest example of a magnetic texture in which the order parameter (magnetization in a FM or staggered magnetization in an AF)  varies along one spatial direction (see Fig.2(b)) separating regions with different orientations of the order parameter. Similar to FMs, the thickness of the antiferromagnetic domain wall is defined by the competition between the magnetic anisotropy and the magnetic stiffness, which arises from exchange. Typically, the domain wall thickness is much larger than the interatomic distances and the domain wall itself can be considered as a smooth texture. Such walls in an AF, in contrast to a FM, carry no macroscopic magnetization. Hence, the rotation of the staggered magnetization within the domain wall can proceed either through the domain wall plane (Bloch-type) or perpendicular to it (N\'eel-type) without additional contribution from the dipole energy.  An exception is provided by a domain wall in a monolayer AF on a heavy-metal substrate. Such a wall can have thicknesses of a few interatomic distances and can show a small net magnetization. \cite{Bode2006}

A magnetic \emph{skyrmion} is a localized, particle-like excitation in which the magnetization (in a FM) or the staggered magnetization (in a collinear AF) is whirling and twisting in all directions (see Fig.2(c)). Stability of the skyrmions is enhanced by their topological properties and thus they are considered as  attractive candidates for transporting information. 

The topological charge of skyrmions is defined similarly for AFs and FMs in terms of the directional order parameter, mapping the staggered magnetization of AFs to the magnetization of FMs. This definition is based on the formal similarity of the equations describing the static skyrmions. However, the properties of ferromagnetic and antiferromagnetic skyrmions are not identical. In particular, the staggered magnetization reverses under the permutation of the magnetic sublattices and the states with  opposite orientation of staggered magnetization are physically indistinguishable. This means that in AFs there is no difference between a skyrmion and an antiskyrmion. \cite{Keesman2016} 

 Formation of skyrmions can be governed by  Dzyaloshinkii-Moriya type interactions (DMI) which are relevant in systems with broken inversion symmetry and  spin-orbit coupling. DMI favors perpendicular orientation of neighboring spins and competes with the collinear orientation favored by the Heisenberg exchange. In a FM, in the continuous limit, DMI yields a contribution to the energy density that is an antisymmetric combination of the magnetization and its spatial derivatives (Lifshitz invariants). \cite{Keesman2015} In an AF, the analogous contribution can involve, depending on the details of the magnetic structure, the combinations of the staggered magnetization, the macroscopic magnetization and their space derivatives. However, the most important terms for the skyrmion formation are the terms containing only staggered magnetization. \cite{Keesman2016}  


Apart from stabilization, another experimentally  challenging problem is the controllable creation of skyrmions. In a FM, skyrmions can be created and manipulated by a magnetic field.
In an AF, which is largely unresponsive to a magnetic field, an alternative way to create antiferromagnetic skyrmions is to pump spin-polarized current \cite{Zhang2015b}  or to introduce impurities in the system. \cite{Raicevic2011}


While the static properties of the textures in FMs and AFs have many similarities, their dynamics is substantially different. The main differences appear from the analysis of the possible driving forces and underlying mechanisms. For the dynamics of  domain walls, driving forces can be divided into two classes: i) those which do not depend on the domain wall structure (so-called ponderomotive forces), and ii) those which do (non-ponderomotive). An example of a ponderomotive force in a FM is given by the external magnetic field applied parallel to the magnetization in one domain. The field removes the degeneracy between the domains, and the {\it ponderomotive force is proportional to the  difference in energy density} of the domains, independent of the nature of the domain wall. Such a force pushes the domain wall toward the energetically less favourable domain.  

{\it Ponderomotive forces -} In an AF, the ponderomotive force can be created by the uniform magnetic field which splits degeneracy of 90$^\circ$ domains or by the staggered field $\mathbf{B}_\mathrm{N}$ we discussed above in the context of the N\'eel spin-orbit torque. In contrast to FM, the efficiency of the uniform magnetic field in moving antiferromagnetic domain walls is low because of the low magnetic susceptibility of AFs. A much more efficient force is generated by the staggered field. \cite{Gomonay2016} Moreover, the staggered field is the only proposed physical field able to split the degeneracy between 180$^\circ$ domains and move the 180$^\circ$ domain walls in AFs.   

The N\'eel spin-orbit torque can also exist in FMs. However, the corresponding force generated by the staggered field produces no effect in FMs and thus is a unique feature of antiferromagnetic dynamics. A similar type of force, which is important for AFs and unimportant for FMs, can be generated by the gradient of a magnetic field. This force originates from the nonzero intrinsic magnetization which appears in the region of inhomogeneity. \cite{Tveten2016} Such a mechanism can be pronounced in synthetic AFs, where the inter-sublattice exchange is weaker compared to crystalline AFs.

{\it Non-ponderomotive forces -} The non-ponderomotive forces can be generated by spin currents which transfer torque directly to the domain wall.  In this case the spin polarization should be directed perpendicular to the plane in which the staggered magnetization rotates within the domain wall. For example, a spin current polarized perpendicular to the domain wall plane effectively pushes a N\'eel wall and produces no effect on a Bloch wall, as was theoretically demonstrated for AFs in Ref. \onlinecite{Shiino2016}.  Various mechanisms of spin current generation in AFs are similar to those used in FMs. For example, a spin current can be pumped from a spin-polarised STM tip, \cite{Wieser2011} due to FMR \cite{Merodio2014b,Frangou2016} or spin Hall effect. \cite{Moriyama2014,Moriyama2015}

 The non-ponderomotive forces in FMs can be also generated by a charge current which is polarized while passing through one of the domain. As they go through the domain wall, the spin-polarized electrons experience a nonuniform magnetization and generate a torque trying to align their spins with the local magnetic moments via the spin-transfer mechanism. An analogous mechanism 
 was also predicted for AFs. \cite{Hals2011,Brataas2015} {In this picture, electrons adiabatically adjust their spin to the space varying quantization axis set by staggered magnetization. The local transversal spin component creates a torque on the domain wall. In another, simple intuitive interpretation, the antiferromagnetic domain wall is viewed as consisting of two coupled ferromagnetic domain walls which form two independent channels for charge current. The spin-polarized electrons are supposed to travel within a given magnetic sublattice, with suppressed hopping between the sublattices.} 


Another source of the forces in AFs, as well as in FMs, are temperature gradients \cite{Selzer2016}  which generate magnon fluxes. In this case the force appears from the transfer of the \emph{linear momentum} to the domain wall. The underlying mechanism can be related to the reflection of circularly polarised magnons from the precessing domain wall \cite{Tveten2014} or to the reduction of the linear momentum of the transmitted magnons due to the interaction with the domain wall (so-called redshift). \cite{Kim2014c}  As different mechanisms are involved in the interaction between magnons and the AF texture, the temperature gradient can create both types of forces.

In contrast to the domain walls, skyrmions embedded within a homogenenous magnetic environment can be driven only by the non-ponderomotive forces stemming, e.g., from 
 the spin transfer torque. The spin torques acting on an antiferromagnetic skyrmion do not create a Magnus force \cite{Zhang2015b,Barker2016,Velkov2016} because two magnetic sublattices experience the Magnus force of  opposite sign. This compensation of partial Magnus forces means that unlike in FMs, the antiferromagnetic skyrmions will not deflect toward sample boundaries where they can disappear. Moreover, their mobility is strongly enhanced \cite{Velkov2016} due to the absence of the gyrotropic force.
 
  Not only spin currents can move an antiferromagnetic skyrmion but, inversely, an antiferromagnetic skyrmion can influence the motion of spin polarized electrons as was demonstrated by the topological spin Hall effect. \cite{Buhl2017} The physics of this effect is related to the complexity of the electronic states in AF.  Unlike in FMs, the electronic states are doubly-degenerate and electrons have more complex, SU(2), gauge structure. The topological spin Hall effect has been proposed for the detection of AF skyrmions. For a more detailed discussion of the topological phenomena we refer to the article on topological antiferromagnetic spintronics in this focused issue.

\section{Spin waves, magnons and associated phenomena}
Magnons, or their classical counterpart, spin waves, play an important role in the transport properties of AFs. They can carry spin current \cite{Wang2014d,Hahn2014,Moriyama2014,Merodio2014b,Frangou2016,Moriyama2015,Lin2016,Rezende2016a,Saglam2016} and contribute to heat flow. \cite{Brataas2015} In addition, they show much reacher physics compared to FMs.

To start with, we consider the main differences between ferromagnetic and antiferromagnetic spin waves which are important for spin transport. In FMs, spin waves always carry a spin. This means that in FMs, a magnon flux is equivalent to  a dc spin current. In AFs, in contrast, spin waves may or may not carry a spin, as oscillations of two coupled sublattices are involved. 

Spin waves can be either linearly or circularly polarized, depending on the magnetic symmetry of the AF. The average spin in the linearly polarized waves is zero and thus is not transmitted. However, the circularly or elliptically polarized waves carry a spin. The spin orientation is opposite for clockwise and counterclockwise rotation. In addition, the modes with opposite spins have the same energy and are equally represented in the equilibrium state. 
This, in particular, explains why the observation of the spin Seebeck effect in AFs (the generation of a spin current by a temperature gradient) requires the presence of an external magnetic field. It either splits the degeneracy of the modes \cite{Wu2016} or induces a nonzero magnetization due to the canting of the sublattice magnetizations in the spin-flop phase. \cite{Seki2015} Without an auxiliary magnetic field the spin Seebeck effect can be observed in compensated ferrimagnets, \cite{Ohnuma2013} in which the staggered magnetization is formed by non-equivalent magnetic atoms and spin wave modes are non-degenerate even in the absence of a magnetic field.

A spin current pumped into an AF modifies the distribution of magnons and can generate a magnon flux with nonzero spin. The ability of AFs to convert electronic spin current into magnonic spin current was demonstrated in the experiments with antiferromagnetic insulators. \cite{Lin2016,Hahn2014,Wang2014d,Moriyama2015,Qiu2016} An antiferromagnetic layer inserted between a spin emitter and a spin detector transmitted spin current and in some cases even resulted in enhancement of the detected signal. This effect was interpreted as originating from magnons, as electronic spin transport through the antiferromagnetic insulator was excluded. Similar effects of spin transmission \cite{Moriyama2014,Merodio2014b,Frangou2016} were also observed in metallic AFs (Fig.2(d)). However, a microscopic interpretation of all the mentioned experiments is still incomplete because of the large number of processes involved in the formation of the detected signals. Theoretical models attribute magnonic spin transport in AFs to magnon diffusion \cite{Rezende2016a} or to the coherent propagation of the evanescent modes. \cite{Khymyn2016,Takei2015}

While spin waves in FMs transfer only the normal spin current, spin waves in AFs can also transfer a staggered spin current which can be produced, e.g., by applying a temperature gradient across an antiferromagnetic insulator. \cite{Brataas2015} This current, which is invisible by standard spin-sensitive techniques, is manifested through the additional contribution of the excited magnons to the heat conductance.

Similar to the neutral bosonic condensates, magnons can also show resistance-free, or superfluid, flow. Superfluidity can occur in isotropic (easy-plane) FMs \cite{Takei2014, Skarsvag2015} or antiferromagnetic  insulators \cite{Takei2014b,Takei2015,Qaiumzadeh2017} with negligible spin-orbit coupling and a very low damping constant.  The analogy between magnon-mediated and real superfluidity is based on the conservation of the magnon number in an axial-symmetric AF and on the degeneracy of states obtained by the coherent rotation of all the magnetic moments. A spin current pumped into the system generates an inhomogeneous distribution of the magnetic moments which transfers spin through the ferromagnetic or antiferromagnetic sample. The value of the spin supercurrent is proportional to the gradient of the rotation angle. Distinctions  between the spin superfluidity in FMs and in AFs appear mainly in the high-temperature regime and need further investigation. \cite{Takei2014b}   

   Coherent spin transport in an AF can be mediated not only by magnons.  In  materials with strong quantum correlations, e.g., quantum spin chains, spin current can be also carried by spinons -- particle-like zero-gap spin excitations over the quantum-correlated ground state. Recently, spinon spin currents were demonstrated by observing a longitudinal spin Seebeck effect in Sr$_2$CuO$_3$. \cite{Hirobe2016} 
    
To conclude, we have considered the key distinctions in dynamics and magnetic textures between FMs and simple collinear AFs with two magnetic sublattices. AFs can also have more complex structures, with non-collinear spins and a larger number of magnetic sublattices. These  systems also show the exchange-enhanced dynamics. The current-induced dynamics and the formation and manipulation of  textures in non-collinear AFs are open research areas which can reveal exciting physics and new sophisticated approaches in antiferromagnetic spintroncis.       
      
%

\end{document}